\documentclass[
    a4paper,
    prb,
    bibnotes,
    twocolumn,
    superscriptaddress,
    groupedaddress,
    floatfix
]{revtex4-1}
\usepackage{graphicx,color}
\usepackage{amssymb,amsmath}
\usepackage{amsmath,float}
\usepackage{epstopdf}
\usepackage[colorlinks=true,citecolor=blue,urlcolor=blue,linkcolor=blue]{hyperref}

\begin{document}
\title{Comparative study of phonon spectrum and thermal expansion of \\graphene, silicene, germanene and blue phosphorene}
\author{Xu-Jin Ge}  
\author{Kai-Lun Yao}  
\author{Jing-Tao L\"u}  
\affiliation{School of Physics and Wuhan National High Magnetic Field Center, Huazhong University of Science and Technology, Wuhan 430074, P. R. China}

\begin{abstract}
Based on first-principles calculation using density functional theory, we study
the vibrational properties and thermal expansion of mono-atomic two-dimensional
honeycomb lattices: graphene, silicene, germanene and blue phosphorene. We
focus on the similarities and differences of their properties, and try to
understand them from their lattice structures. We illustrate that, from graphene
to blue phosphorene, phonon bandgap develops due to large buckling-induced
mixing of the in-plane and out-of-plane phonon modes. This mixing also
influences their thermal properties. Using quasi-harmonic approximation, we
find that all of them show negative thermal expansion at room temperature.
\end{abstract}
	
\pacs{65.80.g, 65.40.De, 63.22.Np}
\maketitle
\section{Introduction}

Since the discovery of graphene\cite{novoselov_electric_2004,zhang_experimental_2005}, two-dimensional (2D) materials, from
mono-atomic single layer silicene \cite{cahangirov_two-_2009,vogt_silicene:_2012,feng_evidence_2012}, germanene\cite{cahangirov_two-_2009}, 
phosphorene\cite{li_black_2014,liu_phosphorene:_2014,PhysRevLett.112.176802,Sun20162098}, 
to transition metal mono- and di-chalcogenides\cite{NovoselovPNAS2005,mak_atomically_2010,splendiani_emerging_2010}, have been the focus of intense
study in the field of physics, chemistry and materials sciences\cite{castro_neto_electronic_2009,zhu_semiconducting_2014,cahangirov_two-_2009,li_black_2014,liu_phosphorene:_2014,geim_van_2013,gupta_recent_2015,vogt_silicene:_2012,feng_evidence_2012,jiang_negative_2014,nika_thermal_2015,nika_two-dimensional_2012,balandin_thermal_2011,yang_how_2012,PhysRevLett.112.176802,Sun20162098,mak_atomically_2010,splendiani_emerging_2010,tao_silicene_2015}. One of the 
most important driving force
for exploring these 2D materials is the possibility of using them to build
next-generation electronics. Thus, their electronic, optical, and magnetic
properties have been extensively studied, experimentally and theoretically.
Among them, silicene has the advantage of being easily incorporated into silicon
based electronics. Very recently, silicene based field effect transistor has been
demonstrated\cite{tao_silicene_2015}.

\begin{figure}[h]
\includegraphics[scale=0.3]{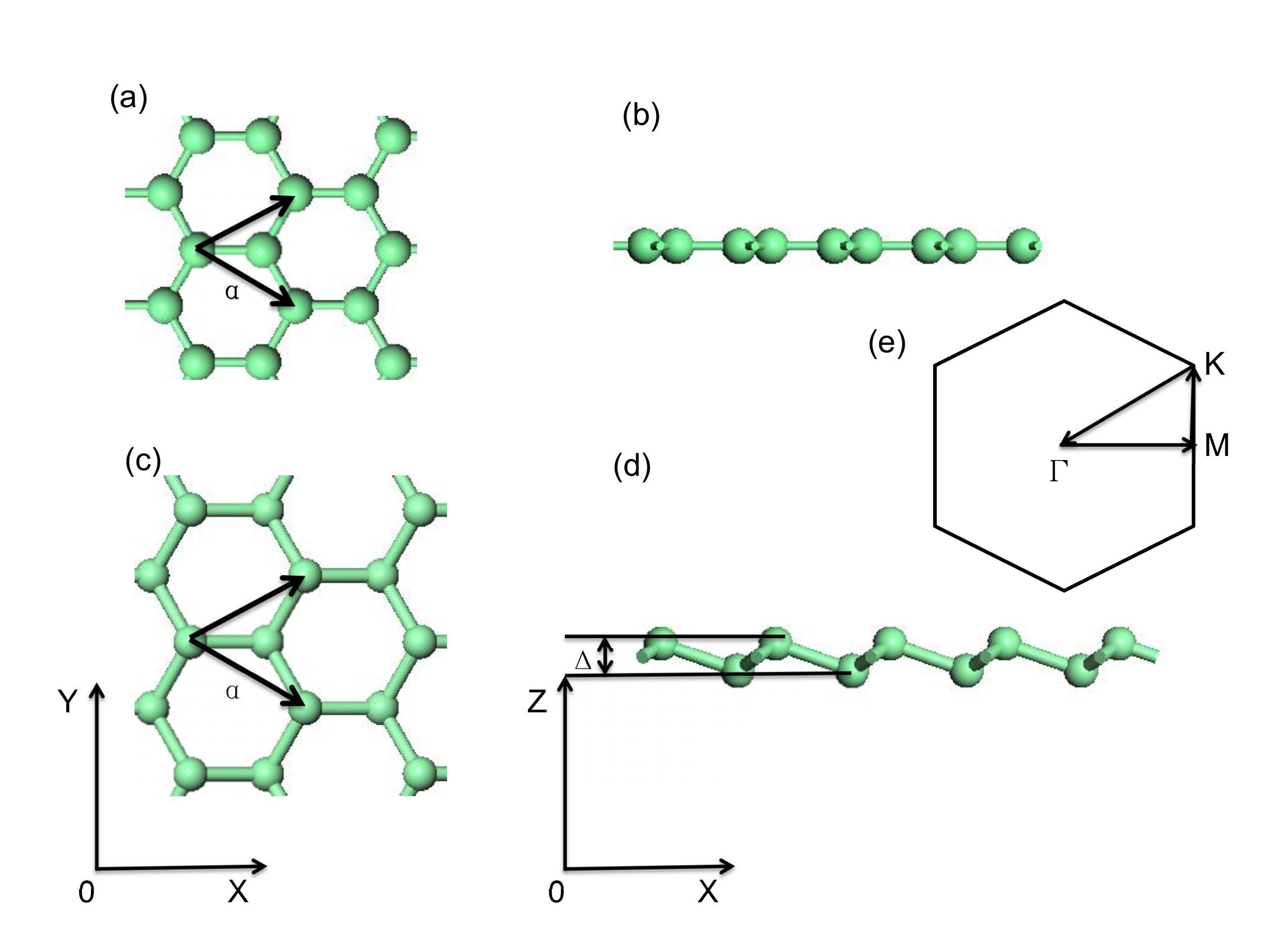}
\caption{Structures of graphene, silicene, germanene and blue phosphorene (a)-(d) and the corresponding first Brillouin zone (e). 
(a) and (c): Top view. (b) and (d): Side view. Graphene is flat (a-b), while all others are buckled (c-d). The lattice constants are $a=2.47, 3.87, 4.06, 3.28$ \AA, and $\Delta = 0, 0.45, 0.69, 1.24$ \AA, for graphene, silicene, germanene.
and blue phosphorene, respectively. }
\label{fig:structure}
\end{figure}

One exciting further direction in the study of 2D materials is to build 2D van
der Waals (VDW) heterostructures by reassembling different kinds of 2D single
layer materials together\cite{geim_van_2013}. Stability of these VDW heterostructures
depends sensitively on their thermal
and vibrational properties, which are relatively less studied\cite{PhysRevB.68.035425,mounet_first-principles_2005,aierken_thermal_2015,gan_large_2015,huang_phonon_2015,huang_correlation_2014,sevik_assessment_2014,cai_lattice_2014}.  Here we focus on
one family of such 2D materials, namely honeycomb graphene, silicene, germanene and blue
phosphorene (Fig.~\ref{fig:structure}) \cite{zhu_semiconducting_2014,aierken_thermal_2015,ding_structural_2015}.  Using graphene as the template system, we perform comparative
study on their vibrational and thermal properties.  We study the relation
between structural and thermal/vibrational properties.  Although some of
these properties have been studied
separately\cite{mounet_first-principles_2005,aierken_thermal_2015,gan_large_2015,huang_phonon_2015,huang_correlation_2014,sevik_assessment_2014,cai_lattice_2014,ding_structural_2015,Jiang09,PhysRevLett.106.135501},
here we focus on a comparative study of their similarities and differences. 

We perform Density Functional Theory (DFT) based calculations of the phonon
spectrum at different lattice constants.  Based on these calculations, using
the quasi-harmonic approximation (QHA), we obtain the Gr\"uneisen
parameters, thermal expansion coefficients and other thermodynamic properties
of these 2D materials. We find that at room temperature, the thermal expansion
coefficients of all these 2D materials are negative. It has already been
experimentally demonstrated that the interaction between graphene and the substrate
can be tuned by utilizing their different thermal expansion coefficients.
We anticipate that similar effect is possible for other 2D materials.

\section{DFT calculation}
Our DFT calculations are performed by using the Vienna $ab$ $initio$ Simulation
Package (VASP) \cite{kresse_efficiency_1996,kresse_efficient_1996}.  It is
based on the projected augmented wave (PAW) method and plane wave basis set.
The Perdew-Burke-Ernzerhof (PBE) version of the generalized gradient
approximation (GGA) is used \cite{perdew_generalized_1996}. We note that the
main conclusions of this work do not depend on the exchange-correlation functionals used.  The energy
cutoffs for graphene, silicene, germanene, and blue phosphorene are 750, 500,
400, 400 eV, respectively.  For the structural relaxations, the Brillouin zone
is sampled using the $\Gamma$ centered scheme with at least $11\times11\times1$
$k$ points. For the vibrational and thermal properties, we need a large unit cell to treat the long range interaction, which is important 
for the long wavelength, low frequency phonons near $\Gamma$, and a dense $k$ point sampling for the high frequency optical phonons. In this work, we have used a supercell of at least $7\times7$, and a $k$-point sampling of $4\times4\times1$.

The mechanical and thermal properties are obtained using Phonopy-QHA script
\cite{togo_first-principles_2008,togo_first_2015}.  Firstly, a series of phonon
spectrum using different lattice constants are calculated. For each lattice
constant $a$, the free energy is obtained from
\begin{eqnarray}
F(a,T) &=& E(a) + \sum_{q,j}\frac{\hbar \omega_{a; q,j}}{2} \nonumber\\
&&+ \frac{1}{\beta}\sum_{q,j}{\rm ln}\left[1-{\rm exp}\left(-\beta\hbar\omega_{a;q,j}\right)\right].
\end{eqnarray}
Here, $E(a)$ is the ground state free energy, $\omega_{a;q,j}$ is the
vibrational frequency corresponding to wavevector $q$, mode $j$, $\hbar$ is the
reduced Planck constant, $\beta=(k_BT)^{-1}$ with $k_B$ the Boltzmann constant,
$T$ the temperature.  A third-order Birch-Murnaghan equation of state is then
used to fit the data points. The equilibrium lattice constants at different $T$
are obtained. The thermal expansion coefficient is defined as
\begin{equation}
\alpha(T) = \frac{\partial {\rm ln}a(T)}{\partial T}.
\end{equation}
$\alpha (T)$ can also be obtained from the mode-dependent Gr\"uneisen parameters
\begin{equation}
	\gamma(q,j) = -\frac{a_0}{\omega_{0;q,j}}\left.\frac{\partial \omega_{a;q,j}}{\partial a}\right|_{a_0}
\end{equation}
as
\begin{equation}
\alpha(T) = -\frac{1}{4V_0 B}\sum_{q,j}c_v(q,j)\gamma(q,j).
	\label{eq:grutheory}
\end{equation}
Here, $c_v$ is the heat capacity at constant volume, $B=-V\partial P/\partial
V$ is the bulk modulus, $a_0$ is the equilibrium lattice constant, $V_0$ is the
equilibrium unit cell volume,  and $\omega_{0;q,j}$ is the corresponding
vibrational frequency. Note that, in our calculation, we fixed the length of
the unit cell in the direction perpendicular to the 2D plane. Thus, we have $4$
instead of $9$ in Eq.~(\ref{eq:grutheory}).

For 2D materials, the ZA mode is very soft, and a slight reduction of the
lattice constant may result in negative phonon frequencies near $\Gamma$ point.
This means the applied strain should be small enough. Otherwise, the QHA is not
valid any more. One important difference between our and previous calculations
is that we have used smaller strain of $\pm 0.5\%$.  Actually, due to this
difference, our results for blue phosphorene are quite different from those of
Ref.~\onlinecite{aierken_thermal_2015}. We have compared results using
different strains to show how sensitively the thermal expansion depends on the
applied strain. To validate our results, we also calculated the thermal
expansion coefficient using the Gr\"uneisen theory from the data points at
strain of $\pm 0.2\%$, with $300\times 300$ $k$-point sampling.  This means
that we ignore the contribution of phonon modes with wavelength larger than
$\sim 0.07 $ $\mu$m. This cutoff is reasonable since in 2D materials ripples 
of similar size form and break the long range order. 


\section{Results}
\begin{center}
\begin{figure}[ht]
\includegraphics[scale=1.04]{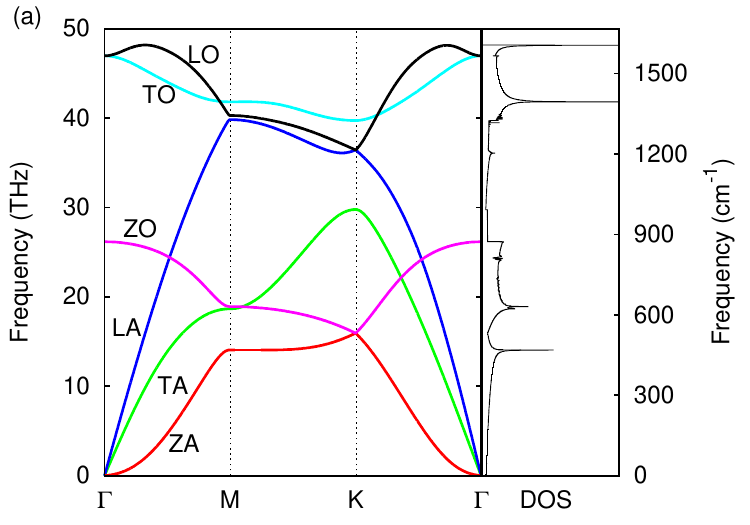}
\includegraphics[scale=1.0]{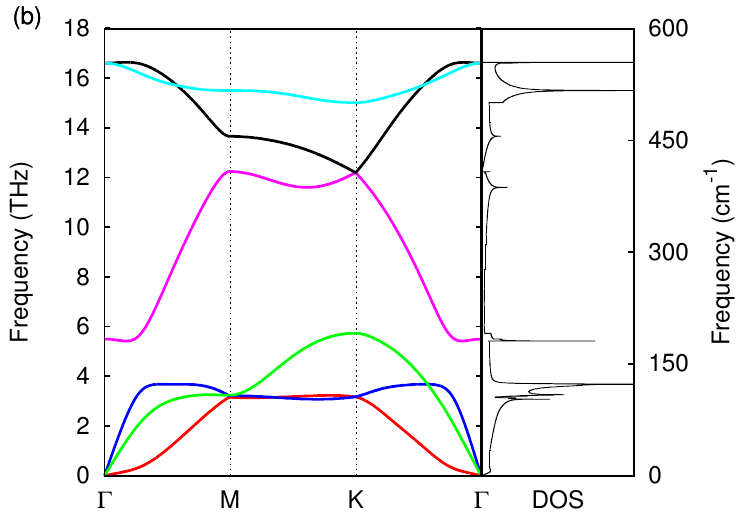}\\
\includegraphics[scale=1.0]{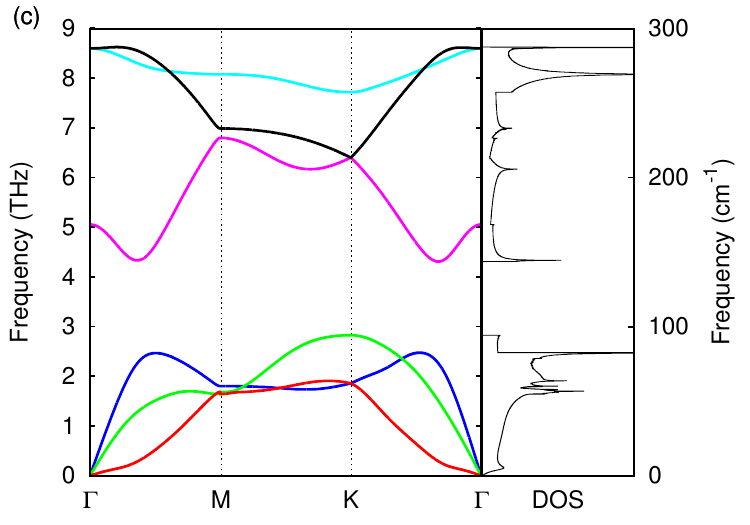}
\includegraphics[scale=1.0]{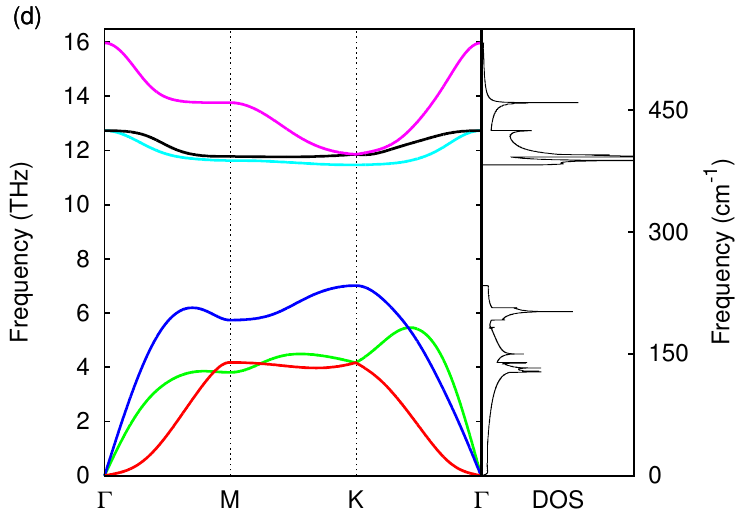}\\
\caption{Phonon dispersion and density of states of graphene (a), silicene (b), germanene (c) and blue phosphorene (d). We use the following color code: LA (blue), TA (Green), ZA (Red), ZO (violet), TO (light blue), LO (black).}
\label{fig:bandos}
\end{figure}
\end{center}

The calculated phonon dispersion relations along high symmetry lines within the
Brillouin zone are shown in Fig.~\ref{fig:bandos} together with the phonon
density of states (DOS). The dispersion lines are similar due to their similar
honeycombed lattice structures. Graphene has a mirror symmetry about the atomic
plane, such that the atomic motions along $Z$ direction are decoupled from
those in the $X$-$Y$ plane in the harmonic approximation.  The acoustic and optical modes along $Z$ direction (ZA (red) and ZO(purple)) do not couple with other phonon modes, resulting in crossings of dispersion lines in graphene. For silicene, germanene and blue phosphorene, the slight buckling of the
atoms in $Z$ direction breaks the mirror symmetry, leading to hybridization of ZA and ZO
with other modes. The crossings turn into avoid-crossings. The
hybridization becomes stronger for larger buckling. This results in (1) the development of
phonon bandgaps, (2) the reduction of phonon group velocity. Both of them reduce effectively the phonon thermal conductivity. Interestingly, the large buckling in blue phosphorene
results in a larger $\Gamma$ point ZO frequency than that of degenerate TO and LO
modes. This does not happen in silicene and germanene.

The buckling of atomic structure does not
change the 3-fold rotational symmetry of the lattice. Due to 
this rotational symmetry, two degenerate points show up at $K$ point in
the dispersion relations of all the materials considered. The quadratic dispersion of ZA mode near $\Gamma$ point in graphene is protected by the mirror and rotational symmetries around $Z$. 
The quadratic dispersion leads to a non-zero DOS at $\omega=0$.
But for other materials, the slightly breaking of mirror symmetry due to buckling introduces a small linear dispersion into the quadratic form.

\begin{center}
\begin{figure}[h]
\includegraphics[scale=1.0]{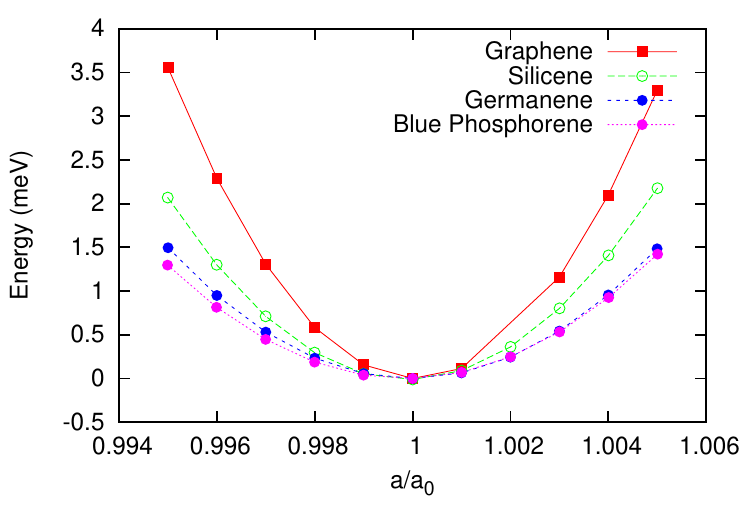}
\includegraphics[scale=1.0]{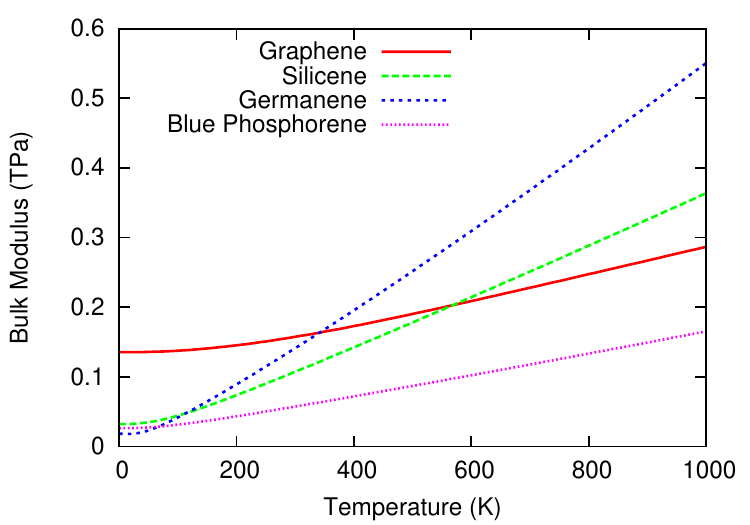}
\caption{(Left) Energy per unit cell as a function of relative lattice constant $a/a_0$　within the range of $-0.005 \le a/a_0 \le 0.005$, with $a_0$ the equilibrium lattice constant. (Right) Temperature dependence of bulk modulus.}
\label{fig:ev}
\end{figure}
\end{center}

To study the thermodynamic properties within the QHA,
we performed a series of calculations by changing the lattice constant within
the range of $\pm 0.5\%$.  The energy-lattice-constant relationship is plotted in
Fig.~\ref{fig:ev}. We can see that the stiffness goes down from graphene to silicene, germanene
and blue phosphorene. Correspondingly, the calculated bulk modulus follows the same trend.

\begin{center}
\begin{figure}[h]
\includegraphics[scale=1.0]{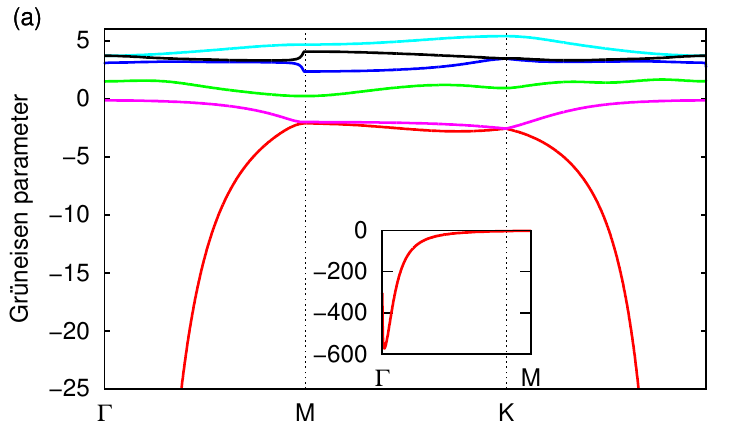}
\includegraphics[scale=1.0]{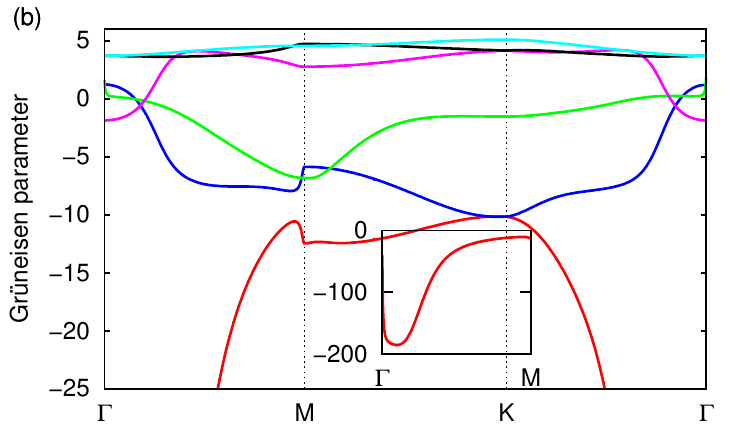}\\
\includegraphics[scale=1.0]{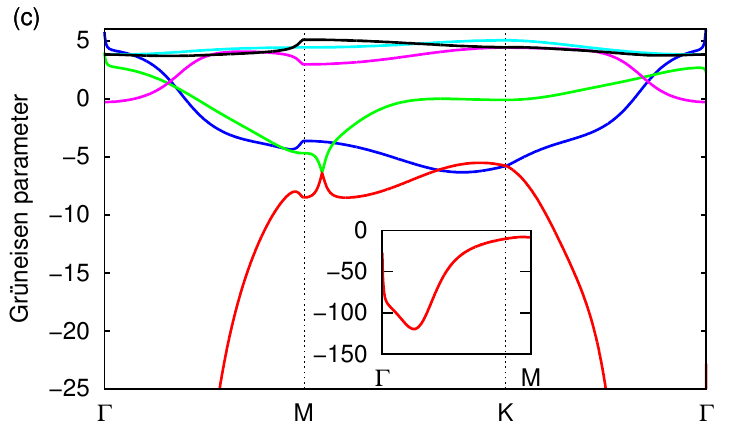}
\includegraphics[scale=1.0]{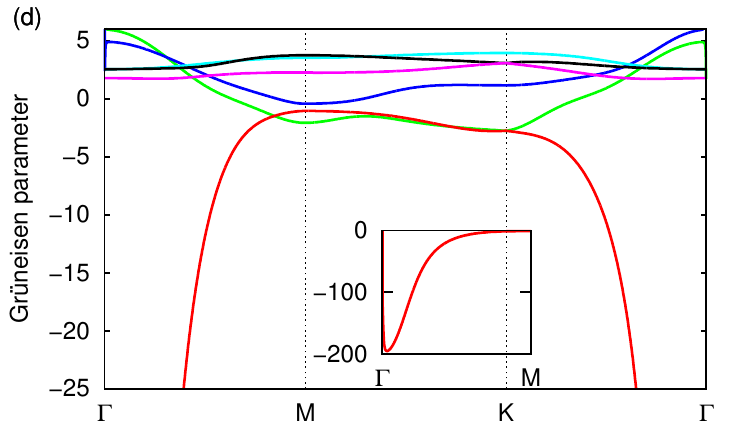}
\caption{Gr\"uneisen parameters of graphene (a), silicene (b), germanene (c)
and blue phosphorene (d) along the high symmetric lines within the Brillouin
zone. Inset: the Gr\"uneisen parameters of ZA mode from $\Gamma$ to $M$ in full
range. We use the same color code as that in Fig.~\ref{fig:bandos}.}
\label{fig:gru}
\end{figure}
\end{center}

Using the phonon dispersion at $a=0.998a_0, a_0, 1.002a_0$, we calculated the
mode Gr\"uneisen parameters as shown in Fig.~\ref{fig:gru}. In
Table.~\ref{table:phonon1}, we compare our results with previous works,
especially with those of Ref.~\onlinecite{mounet_first-principles_2005}.  They
show reasonable agreement. This comparison validates the calculation procedure
we used here for other materials.  We can try to understand the results
starting from graphene. As has already been shown by many previous
works\cite{mounet_first-principles_2005,aierken_thermal_2015,huang_phonon_2015,huang_correlation_2014,sevik_assessment_2014},
the graphene ZA and ZO modes have negative Gr\"uneisen parameters, explained by
Lifshitz\cite{lifshitz1952thermal}.  This abnormal hardening of phonon modes
upon expansion is a general feature of the 2D out-of-plane mode, and the reason
why graphene shows negative thermal expansion. All other modes with in-plane
motion have normal, positive Gr\"uneisen parameters. As we have mentioned,
there is no coupling between ZA, ZO modes with modes in the $X$-$Y$ plane.
There is a clear distinction of these modes in the calculated Gr\"uneisen
parameters. For silicene, germanene, and blue phosphorene, due to the buckling,
atomic motions in $Z$ and $X$-$Y$ directions mix. Away from the $\Gamma$ point,
there are more modes with negative Gr\"uneisen parameters. But independent on
elements, the TO and LO modes have Gr\"uneisen parameters around $2$ . Finally,
one notices that due to large buckling in blue phosphorene, the ZO mode at
$\Gamma$ point has a larger frequency than the LO and TO modes, and a positive
Gr\"uneisen parameter, contrary to the other three materials.  This shows a
gradual lost of 2D properties of the ZO mode.

\onecolumngrid

\begin{center}
\begin{figure}[htp]
\includegraphics[scale=0.5]{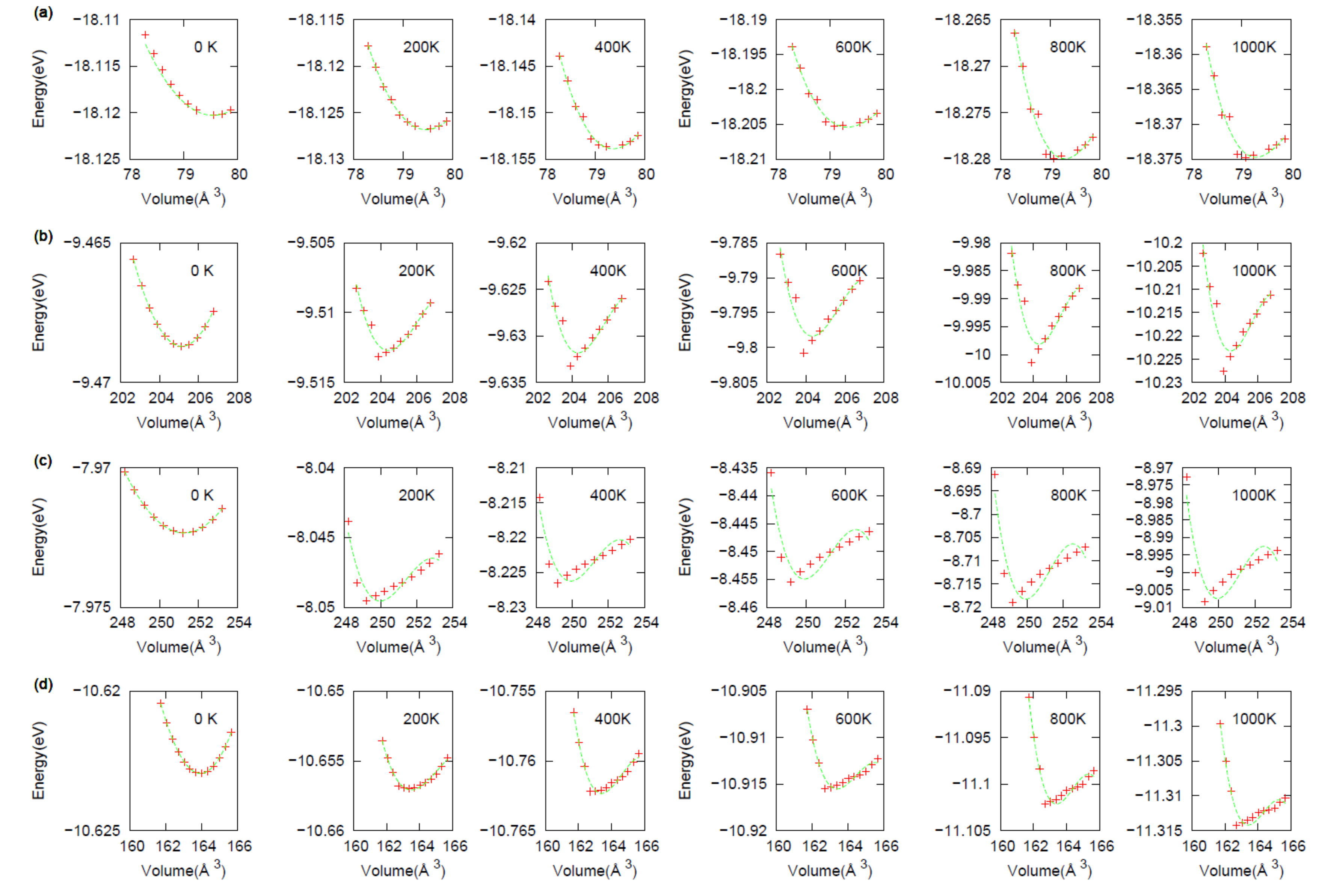}
\caption{Fitting (Green dashed line) of the total free energy (electronic plus phononic) to the 3rd order Birch-Murnaghan equation of state at representive temperatures for graphene (a), silicene (b), germanene (c) and blue phosphorene (d). }
\label{fig:bmfitting}
\end{figure}
\end{center}

\begin{center}
\begin{figure}[htp]
\includegraphics[scale=1.0]{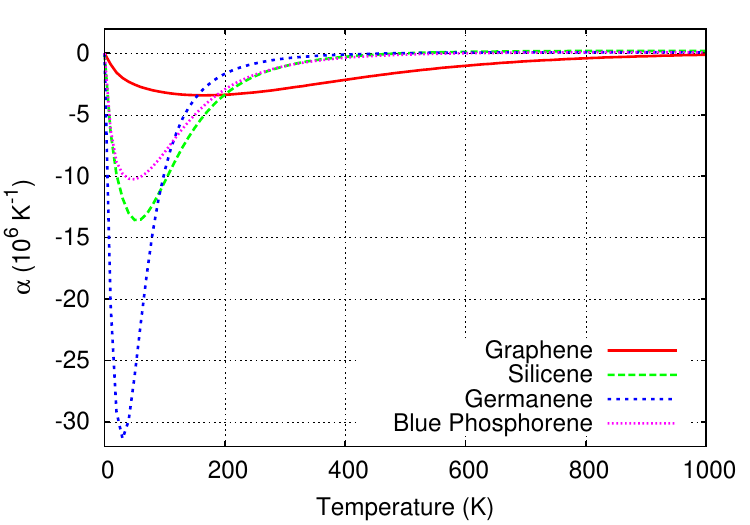}
\includegraphics[scale=1.0]{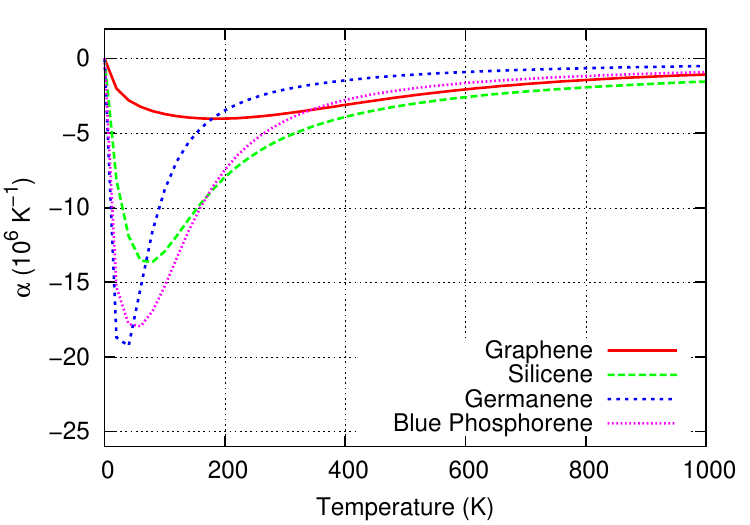}
\caption{Thermal expansion coefficient as a function of temperature. (Left) Fitting the third-order Birch-Murnaghan equation of state. The strain applied are $\pm 0.5\%$. (Right) Calculated from the Gr\"uneisen method.}
\label{fig:te}
\end{figure}
\end{center}
\twocolumngrid

\onecolumngrid
\begin{center}
\begin{table}[htp]
\begin{tabular}{ccccc}
\hline
\hline
Graphene & this work & GGA & LDA & experiment\\  
\hline
$\omega_{ZO}$&873&881\cite{mounet_first-principles_2005},884\cite{Wirtz2004141}&890\cite{PhysRevB.67.035401}, 893\cite{Wirtz2004141},884\cite{PhysRevB.80.033406} &\\
\hline
$\omega_{TO/LO}$ & 1566&1554\cite{mounet_first-principles_2005},1569\cite{Wirtz2004141}&1586\cite{PhysRevB.77.125401},1595\cite{PhysRevB.67.035401},1597\cite{Wirtz2004141},1560\cite{PhysRevB.80.033406} &\\
\hline
$\gamma_{TA}$& 0.6&0.8\cite{mounet_first-principles_2005}&0.8\cite{PhysRevB.80.033406} \\
\hline
$\gamma_{LA}$ & 1.4&1.7\cite{mounet_first-principles_2005}&1.6\cite{PhysRevB.80.033406} \\
\hline
$\gamma_{ZO}$ & -0.1 & 0\cite{mounet_first-principles_2005}&-0.1\cite{PhysRevB.80.033406}\\
\hline
$\gamma_{LO/TO}$ & 1.9&1.86\cite{PhysRevB.83.115449},1.9\cite{mounet_first-principles_2005}&1.9\cite{PhysRevB.80.033406}&1.99\cite{PhysRevB.79.205433},1.80\cite{ding_stretchable_2010}\\ 
\hline
TEC(10$^{-6}$K$^{-1}$)& -2.9$^{a}$, -3.7$^{b}$&-3.7\cite{mounet_first-principles_2005}&&-7\cite{bao_controlled_2009},-8\cite{yoon_negative_2011},-7\cite{0957-4484-21-16-165204}\\
\hline
\end{tabular}
\caption{Comparison of graphene phonon frequencies $\omega$ , Gr\"uneisen parameters $\gamma$  at $\Gamma$ point and thermal expansion coefficient (TEC) at room temperature to previous works. Here, $a$ means results from QHA, and $b$ means that from  Gr\"uneisen theory.}
\label{table:phonon1}
\end{table}
\end{center}
\twocolumngrid
\begin{center}
\begin{table}
\begin{tabular}{ccccccc}
\hline	
\hline	
TEC($10^{-6}$ K$^{-1})$ & silicene & germanene & blue phosphorene\\
\hline
this work &-1.0$^{a}$,-5.3$^{b}$ &-0.43$^a$, -2.1$^b$ & -1.0$^a$, -4.2$^b$\\
\hline
previous works &-7.2\cite{huang_phonon_2015}& -2.4\cite{huang_phonon_2015}&7.8\cite{aierken_thermal_2015},-0.5\cite{Sun20162098}\\
\hline
\end{tabular}
\label{table:tec}
\caption{Comparison of room temperature thermal expansion coefficient (TEC) obtained here to
previous calculations. }
\end{table}
\end{center}


From the series of calculations, we can obtain the thermal expansion
coefficients as a function of temperature using two methods. The left panel of
Fig.~\ref{fig:te} shows results from fitting the third-order Birch-Murnaghan
equation of state, while the right panel shows that from the Gr\"uneisen theory. 
The details of the fitting to the equation of state at representive temperatures are shown in Fig.~\ref{fig:bmfitting}.

The general trends for all the four materials are the same: $\alpha$
starts from zero, goes down and reaches a minimum value. Afterwards, it goes up
monotonically. This can be understood as follows. At low temperatures, the ZA
mode populates much more than all other modes, and it has a large DOS. Thus, it
dominates over other modes and contributes to negative thermal expansion due to
its negative Gr\"uneisen parameter. The ZA mode keeps dominating until certain
temperature. After that, the modes with positive Gr\"uneisen parameters get
populated, and become important, consequently $\alpha$ goes up.  The
temperature at which $\alpha$ reaches its minimum is related to the temperature
at which the heat capacity of ZA modes saturates to its classical value
(Eq.~\ref{eq:grutheory}). The heavier the elements, the lower this temperature.

\begin{figure}[ht]
\includegraphics[scale=0.9]{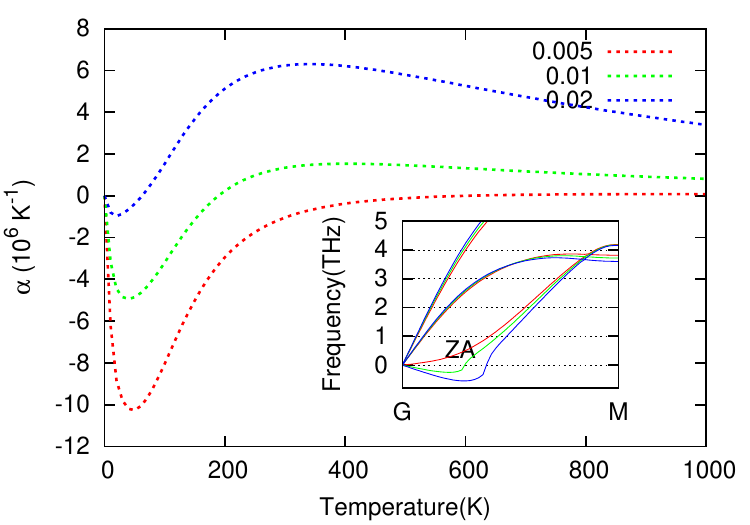}
\caption{Thermal expansion coefficients of blue phosphorene for different
strains from $\pm 0.5\%$ to $\pm 2\%$. This shows that the results depend sensitively on the
strains applied. Inset: The phonon dispersion near $\Gamma$ point at different strain applied. For larger strain, a large fraction of ZA modes goes negative,
indicating the failure of QHA. A small strain ($\pm 0.5\%$) is needed in the
calculation to minimize this effect. }
\label{fig:strain}
\end{figure}

We note that the mode
Gr\"uneisen parameter and consequently the thermal expansion coefficient are minimal change 
of mode frequency as a function of lattice constant and lattice constant as a function of temperature,
respectively. Both are very sensitive to the calculation parameters and approximations used. 
Although our results from the two methods follow similar trends, they are
different quantitatively. 
Actually, the results from fitting the equation of
state depend sensitively on the range of strain applied to the material. 
For 2D materials, the
ZA mode is soft near $\Gamma$ point.  A slight compression of the lattice
constant results in decrease of the phonon frequency. In practical
calculations, modes near $\Gamma$ point go negative, indicating the structure
is not stable (Fig.~\ref{fig:strain} inset), or the QHA used here is not
valid anymore. To minimize this
technical problem, one should keep the strain as small as possible.  
This is why a small strain of $\pm 0.5\%$ was chosen in this work. But, on the
other hand, to fit the results to an equation of state, we need to have data
points span in a reasonably large range of energy. Due to this difficulty, we
argue that it is more appropriate to use the Gr\"uneisen theory to predict the
thermal expansion of 2D materials, as shown in the right panel of Fig.~\ref{fig:te}.

We have also compared the thermal expansion coefficients obtained here with
previous results  from DFT in Table~\ref{table:phonon1} and \ref{table:tec}.
For graphene, we get similar results with
Ref.~\onlinecite{mounet_first-principles_2005}.  For silicene and germanene,
due to the different long wavelength cutoff used, and different $k$-point
sampling, our results are similar to, but quantitatively different from those
of Ref.~\onlinecite{huang_phonon_2015}.  This discrepancy is acceptable.  For
blue phosphorene, we get negative thermal expansion coefficient of $-1.0\times
10^{-6}$ K$^{-1}$ by fitting the equation of state, reasonably agree with
$-0.5\times 10^{-6}$ K$^{-1}$ in Ref.~\onlinecite{Sun20162098}. However, in
Ref.~\onlinecite{aierken_thermal_2015} the authors get a positive value of
$7.8\times 10^{-6} $ K$^{-1}$, much larger than ours.  We believe that the
large discrepancy comes from different range of strain applied. We argue that
too large strain drives the system out of the valid range of QHA
(Fig.~\ref{fig:strain}).

\section{Conclusions and Remarks}
We have studied the vibrational and thermal properties of graphene, silicene,
germanene and blue phosphorene using first-principles calculations based on
QHA.  We have shown that the similarities and differences of their vibrational
and thermal properties can be traced back to their structures. We find that all
the materials considered show negative thermal expansion at room temperature.
Our findings are useful in the design of VDW heterostructures, where different
2D materials are vertically stacked together.  Finally, from the numerical
point of view, we argue that, the calculated thermal expansion coefficients
depend sensitively on the strain applied to the material due to the soft ZA
mode of the 2D materials. Thus, it is more appropriate to use the Gr\"uneisen
theory to study thermal expansion in 2D materials. Meanwhile, more advanced
method going beyond the QHA is needed for more accurate prediction of thermal
expansion coefficient in these 2D materials.  Molecular dynamics simulation can
in principle take into account the full anharmonic interactions, and serves as
a possible solution to the problem. But the
computational cost is huge in order to get accurate thermal expansion
coefficient. We are aware of only one work using this approach\cite{PhysRevLett.106.135501}.

It is worth mentioning that, here all the calculations are done for single
layer without including the substrate. For supported monolayer, the interaction
between the layer and the substrate removes the translational invariance of the
monolayer. $\Gamma$ point frequencies of all modes become nonzero. The absolute
values of the negative Gr\"uneisen parameters becomes smaller. As a result, the
thermal expansion at room temperature becomes less negative or even positive. 
The effect of the substrate on the thermal expansion of graphene nanoribbon has been studied 
in Ref.~\onlinecite{JiangPRB09,PhysRevLett.106.135501}. One should keep
this fact in mind when comparing theoretical to experimental results.

For bulk materials, using Klemens model\cite{Klemens19581,NikaPRB09,NikaAPL09},
the phonon thermal conductivity can be estimated from the dispersion and
Gr\"uneisen parameters obtained here. For example, the phonon group velocity
and density of states can be readily deduced from the dispersion relation, and
the anharmonic interaction between different modes can be estimated from the
Gr\"uneisen parameters. But, for 2D materials, there are subtleties which make
the estimation inaccurate.  Currently, there is still ongoing debate on the
size dependence of phonon thermal conductivity of 2D
materials\cite{LiRMP2012,XuNC2014}. Interesting hydrodynamic phonon transport
is predicted in 2D materials\cite{CepellottiNC2015,LeeNC2015}. All these make 
it difficult to estimate the thermal conductivity from quantities calculated in this work.

\acknowledgements
We thank Nuo Yang and Zelin Jin for discussions. This work was supported by the
National Natural Science Foundation of China (Grant Nos. 11304107, 61371015,
11274130). The authors thank the National Supercomputing Center in Shanghai for
providing help in computations.

\bibliography{2DVibThermo}

\end{document}